\documentclass[11pt]{article}
\usepackage{epsfig}
\usepackage{amssymb}
\usepackage{graphicx} 
\usepackage{subfig} 
\usepackage{color} 

\graphicspath{{immagini/}}

\newcommand{\Didasc}{\itshape}

\author{
L.\ S.\ Di Mauro\thanks{Department\ of Physics and Astronomy, University\ of Catania and INFN Sezione di Catania, Italy; letizia.dimauro@ct.infn.it}$\:$,
A.\ Pluchino\thanks{Department\ of Physics and Astronomy, University\ of Catania and INFN Sezione di Catania, Italy; alessandro.pluchino@ct.infn.it}$\:$,
A.\ E.\ Biondo\thanks{Dept.\ of Economics and Business, Univ.\ of Catania, Italy; ae.biondo@unict.it}}

\title{\bf A Game of Tax Evasion: \\ evidences from an agent-based model} 

\begin{document}
\date{ }
\maketitle

\begin{abstract}
This paper presents a simple agent-based model of an economic system, populated by agents playing different games according to their different view about social cohesion and tax payment. After a first set of simulations, correctly replicating results of existing literature, a wider analysis is presented in order to study the effects of a dynamic-adaptation rule, in which citizens may possibly decide to modify their individual tax compliance according to individual criteria, such as, the strength of their ethical commitment, the satisfaction gained by consumption of the public good and the perceived opinion of neighbors. Results show the presence of thresholds levels in the composition of society - between taxpayers and evaders - which explain the extent of damages deriving from tax evasion.
\end{abstract}


\section{Introduction}

Tax evasion is quite an age-old phenomenon that has been studied for decades, both theoretically and empirically. It can be described as the ``illegal and intentional actions taken by individuals to reduce their legally due tax obligations'' (Alm 2012, p.55). Tax evasion is a damage to the socio-economical environment that deprives governments of their fiscal resources and plays an important role in reducing well-being of societies. The well-known \textit{free rider} problem rises when a selfish citizen consumes public goods and services without properly contributing to related costs, causing inefficiency and bad allocations of governments expenditures for healthcare, education, defence, social security, transportation, infrastructure, science and technology, as widely documented in a vast literature, among which Andreoni \textit{et al.} (1998), Slemrod and Yitzhaki (2002), Torgler (2002), Kirchler (2007), Slemrod (2007). Tax evasion is, also, related to social inequality, as underlined by part of the literature, among which, Alstadsaeter \textit{et al.}  (2017), Bertotti and Modanese (2014, 2016), dealing with the differentiation of the propensity to evade with respect to income and with redistributive aspects. Finally, it matters in terms of social justice, since it specially afflicts poorer people, who do not have the possibility to substitute public services with private ones available in the market at higher prices.
\\In his \textit{An Essay on the Nature and Significance of Economic Science}, Lionel Robbins (1932, p.15) wrote that ``economics is the science which studies human behaviour as a relationship between ends and scarce means which have alternative uses''. Such a very well-known definition, focuses the point that among all routes between ends and means, each person chooses the one that maximally conforms to her personality. A very intriguing corollary would discuss all decisions in terms of their ethical acceptability. Unfortunately, the moral evaluation of chosen objectives is not pertinent of the economic analysis and remains within the attributions of personal free will. More precisely, the individual propensity between cooperation and competition is determinant in setting the deliberated conduct. While the epistemological analysis of such aspects would go far beyond the goal of this paper, the model here presented aims to show the collective relevance of such behavioural elements in driving the decision of each citizen, which reflects also the perceived quality of the public good and the relational feedback received by her surrounding social environment.
\\The initial stages of the formal analysis of tax evasion can be dated back to the Seventies, with contributions by Allingham and Sandmo (1972) and Srinivasan (1973). Despite many similarities, such contributions, which are a propagation of an earlier approach advanced by Becker (1968), differ from each other with respect to optimization procedures, taxpayer’s risk attitudes (which affect second order conditions of chosen objective functions), decision variables, audit probabilities, tax tariffs, and penalty functions. In particular, Allingham and Sandmo (1972), find that income understatement is  decreasing in audit probabilities or in the fine, whereas the dependence on tax rate is more controversial, reflecting income and substitution effects. Yitzhaki (1974) obtained a conter-intuitive result by modelling fines computed on the basis of evaded taxes (instead of the understated income): differently from the empirical evidence shown in Clotfelter (1983), Crane and Nourzad (1987), Poterba (1987), as the tax rate increases, the evasion decreases. Many other studies have been done in the attempt to find a positive relatioship between tax rate and evasion (see for example, among others, Yitzhaki (1987), Panades (2004), Dalamagas (2011), Yaniv (2013)).
\\Such a standard theoretical framework inspired several contributions in related literature, concerning tax evasion and related issues, such as the shadow economy, as in Buehn and Schneider (2012), psychological perception and society (social norms and moral sentiments like guilt or shame), as in Myles and Naylor (1996), Traxler (2006), Fortin \textit{et al}. (2007), Kirchler (2007) and many others. However, tax evasion has been discussed also in contributions based on an econophysics approach, since papers by Lima and Zacklan (2008) and Zaklan \textit{et al}. (2008) where tax declaration/evasion correspond to the two states of spins in the Ising model (1925) of ferromagnetism. More generally, a growing stream of literature presenting agent-based models dealing with tax evasion exists. A survey of such papers could be gained by the joint reading of Bloomquist (2006), Alm (2012), Hokamp (2013), Pickhardt and Prinz (2014), Oates (2015), Bazart \textit{et al}. (2016). The advantage of agent-based models is that they are prone to describe the complexity of aggregate contexts, as documented in previous studies of socio-economic analysis, Pluchino \textit{et al}. (2010, 2011, 2018),  Biondo \textit{et al}. (2013a, 2013b, 2013c, 2014, 2015, 2017). Simulative models can help investigating relevant questions, as the correspondence between the provision of the public good and tax evasion, as in Hokamp (2013), the importance of social norms and auditing, as in Hokamp and Pickhardt (2010), and the effect of social networks on the tax compliance, as in Vale (2015). Such aspects, like many others, can be explained in terms of behavioral attributes, seeking for the roots of decisions in the evolution of personal traits, influenced by the surrounding environment.
\\As reported by the IRS (2016), given the extent of the tax evasion, the expenditures paid by governmental authorities to induce virtuous behaviours are significant. Nonetheless, in many cases, free riders remain unpunished. Honest citizens considering the participation to social costs as a moral imperative are the sole fully compliant taxpayers. We rephrase the provoking question asked by Alm \textit{et al.} (1992): why should people pay taxes? 
\\In order to answer the question, the main motivation of this paper is to combine the agent-based approach to the problem of tax evasion with a flavour of game theory. The free rider problem is one of the most classical example of failures of coordination mechanisms. When people try to obtain the best outcome for themselves, the result might be that, considering the final collective effect, everyone gets the worst instead. In other words, we assist to a conflict between \textit{individual} and \textit{collective} rationality (Rapoport 1974). This creates the paradoxical outcome, know in the game-theory literature as the \textit{prisoner's dilemma} where, despite two persons apply their dominant strategies, they reach a sub-optimal equilibrium. Such a social dilemma motivated a vast amount of literature, regarding the production of public goods, as in Heckathorn (1996), the emergence of social norms and social interaction, as in Hardin (1995), and Voss (2001), among many others.
\\The game-theoretic approach re-defines the meaning of the above-asked question: why should individuals ever decide to cooperate if they have incentives to pursue, firstly, their own self-interest? Different reasons can be reported: first of all, because of altruism, as recalled, for just some examples, in Stevens (2018), Epstein (1993), and Zappal\`a \textit{et al.} (2014); secondly, because of imitation, as in Callen and Shapero (1974), in Elsenbroich and Gilbert (2014) and McDonald and Crandall (2015); finally, because of an assessment of the quality of the public good (i.e. quality of Institutions), as in Nicolaides (2014), La Porta \textit{et al.} (1999), Feld and Frey (2007) and Torgler and Schneider (2009).
\\The present model will try to analyze the aggregate dynamics of a community in which agents decide how to behave, in terms of tax compliance, according to highlighted factors, such as the ethical orientation, the appreciation of the public good and the imitation. The first, very simple, setting of the model will be shown to be able to replicate same results obtained by Elster (1989), who addresses the simplest case of the many-person prisoner's dilemma. Several successive modifications of the same model will be used to validate policy hypotheses and social implications. The paper is organized as follows: in Section 2 we present our Tax Evasion Model; in Section 3 we analyze the impact of a fraction of taxpayers in the evolution of the system, by emphasizing the role of cooperative altruism of tax-payers; in Section 4 we introduce the possibility for the agents to change their behaviors because of two mechanisms (i.e., imitation and assessment of the satisfaction from the public good); in Section 5 we study the effect of three policy parameters regulating the tax rate, the penalty and the audit probability.

\begin{figure}[tbp]
\centering
\includegraphics[width=0.8\columnwidth]{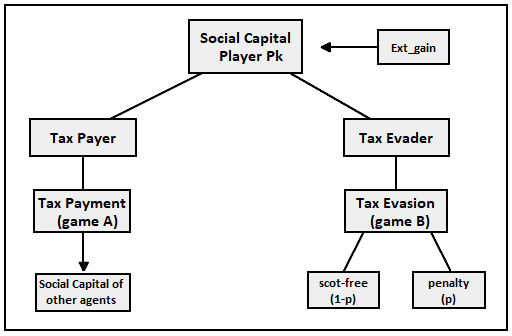}
\caption{{\bf Tax Model}. \Didasc Rules of a minimal Tax evasion model, for a community of $N$ players. Tax evaders play only game B, while taxpayers play only game A. At each turn their individual capital increases through an external gain. }
\label{Tax_Model_bordo2}
\end{figure}
Time $t=0,1,2,\ldots,T$ is a discrete variable indicating the number of played turns. 

\section{Tax Evasion Model}

Consider a community where $N$ players $\{P_i\}_{i=1,2,\ldots,N}$, all endowed with the same initial amount of capital $C_i$, can play one of two very simple games:
\begin{itemize}
\item \emph{Game~A}: at each turn, the player gives away $d$ units of capital, which go to reinforce the capital of other randomly chosen players (1 unit for each);
\item \emph{Game~B}: at each turn, the player loses $h$ units of capital (which will {\it not} be redistributed to other players) with probability $p$, otherwise she does not pay anything.\end{itemize}

Parameters $d$ and $h$ are integer constants, with $h>d$.
From an individually point of view, game~A is of course a losing game, since the player loses some units of capital donating them to other players. On the other hand, from a collective point of view, it may also be regarded as an altruistic behavior, since the player sacrifices her personal wealth to favor other people. From this perspective, playing game~B can be considered as a selfish behavior, since the player prefers to risk paying a penalty just to have a chance of preserving her capital. Furthermore, if she loses, the penalty is greater of the altruistic donation of \emph{Game~A} and the loss does not increase the capital of other players. 

For our purposes, if one looks at the individual capital as the comprehensive monetary value of goods and services (both private and public) enjoyed by each player at the current turn, it is possible to consider the altruistic conduct of the above-sketched metaphoric game as the tax payment, i.e., the choice of a person to reduce her own capital in order to increase someone's else endowment, whereas the selfish behavior stands for tax evasion. In other words, depending on the chosen game, agents are partitioned in two categories, as depicted in Figure~\ref{Tax_Model_bordo2}:
\begin{itemize}
\item \emph{Taxpayers}: altruistic players, indicated as $\{A_i\}_{i=1,2,\ldots,N_{\mathrm a}}$, who always pay taxes playing game~A;
\item \emph{Tax evaders}: selfish players, indicated as $\{S_i\}_{i=1,2,\ldots,N_{\mathrm s}}$, who always evade taxes playing game~B.
\end{itemize}
where $N=N_{\mathrm a} + N_{\mathrm s}$. The initial amount of social capital $C_i(0)=0$ $(i=1,2,\ldots,N)$ evolves in time according to the chosen game.
\\Tax's payment does occur when altruists play game~A, in the sense that we interpreted the donation of $d$ units of capital towards other $d$ randomly chosen players as a taxation. Thus, being the profits gained from taxes equally distributed among the population, they can be interpreted as public services. On the other hand, if people who do not pay taxes (the ones who play game~B) win the game, they get off scot-free and their capital remains the same; losers, instead, are forced to pay a penalty $h$ higher than the tax. The probability $p$ of losing the game is the probability to fall into an audit. Finally, at the beginning of each turn, people's capital is incremented by a certain amount $g<d$ of units, which represents the only external source of gain for the whole community (see again Figure~\ref{Tax_Model_bordo2}). In terms of benefits for any agent, this latter rule allows to consider the social capital $C_i$ as composed by both the accumulated external gain and the value of the enjoyed public services.   
\\The null settings for the first part of the paper is: $p=0.4$ (audit probability), $d=2$ units (tax payment), $h=3$ units (evasion penalty) and $g=1$ units (external gain). Such values will be later changed to show their incidence on results.

\begin{figure}[tbp]
\centering
\includegraphics[width=\columnwidth]{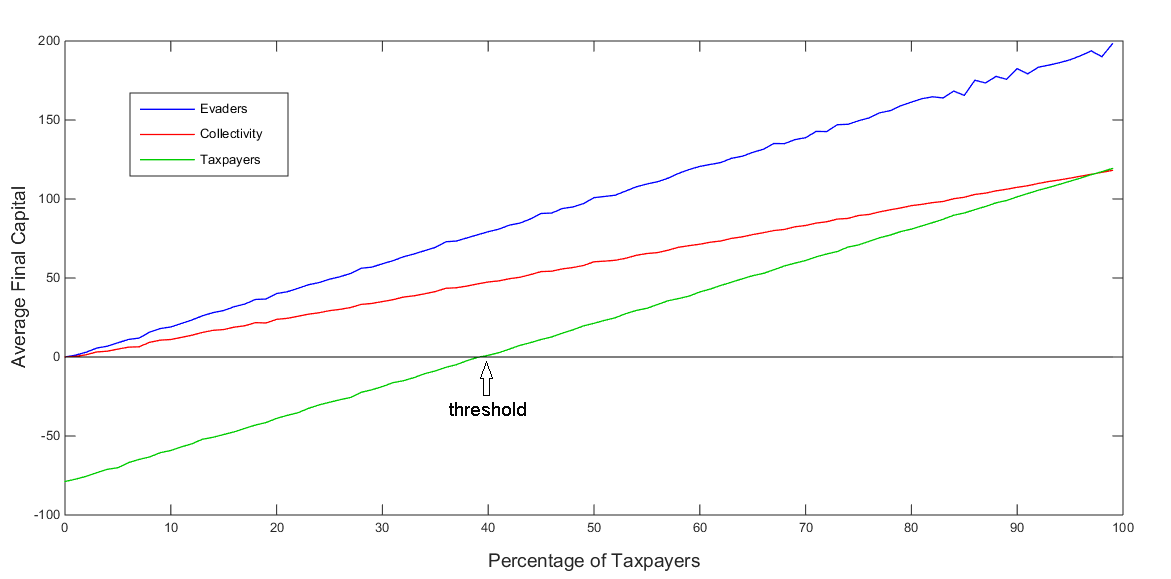}
\caption{{\bf Tax Model}. \Didasc Average final capital $\bar{C}(T)$ (red line) over $T=100$ turns as a function of the percentage $f$ of taxpayers. The average final capitals $\bar{C_a}(T)$ (green line) and $\bar{C_s}(T)$ (blue line) for the two categories of players, taxpayers and tax evaders, are also reported and compared the one with the other. The percentage $f$ varies from 0 to 100 \% with steps of 1 \%. All the capital values have been rescaled in order to have $\bar{C}(T)=0$ for $f=0\%$.}  
\label{capital_vs_perc_mat1}
\end{figure}

\section{Impact of a varying fraction of altruistic players}

In this section we want to investigate the asymptotic behavior of the average social capital $\bar{C}(T)=\frac{1}{N} \sum_{i=1}^N C_i(T)$, calculated at the end of single run simulations with $T=100$ turns per players, in correspondence of an increasing percentage $f$ of taxpayers present in the community (with $f$ varying from 0 to 100 \% at steps of 1 \%).
\\In Figure~\ref{capital_vs_perc_mat1}  the average final capital $\bar{C}(T)$ is reported (red line) as function of $f$, along with its two components $\bar{C_a}(T)=\frac{1}{N_a} \sum_{i=1}^{N_a} C_i(T)$ (green line) and $\bar{C_s}(T)=\frac{1}{N_s} \sum_{i=1}^{N_s} C_i(T)$ (blue line) calculated separately for the altruistic taxpayers and selfish evaders (a horizontal black line at $\bar{C}=0$ is also drawn for comparison).   
The special case $f = 0 \%$ (i.e. only selfish players, $N_{s} = N$) is equivalent to playing only game~B, whereas the opposite case $f = 100 \%$ (i.e. only altruistic players, $N_{a} = N$) corresponds to the execution of game~A only. The three values of social capital have been rescaled in order to have $\bar{C}(T)=0$ when $f = 0 \%$, since -- by definition -- if no one pays taxes the collectivity must have zero benefits. 
\\Looking to the graph, one can see that the average capital $\bar{C_s}(T)$ of the evaders rapidly grows with $f$. This happens because, for increasing values of $f$, there are less and less selfish players surrounded by a growing number of altruistic ones: these players, besides the external gain, enjoy the public services ensured by taxpayers without giving any personal contribution, so their average social capital tends to remain always positive. On the other hand, even the capital $\bar{C_a}(T)$ of taxpayers grows with $f$ since, when there are many altruists who randomly donate units of capital, a large fraction of the donations statistically goes to other altruists, so the altruistic component has, collectively, a small loss of capital.
\begin{figure}[tbp]
\centering
\includegraphics[width=0.6\columnwidth]{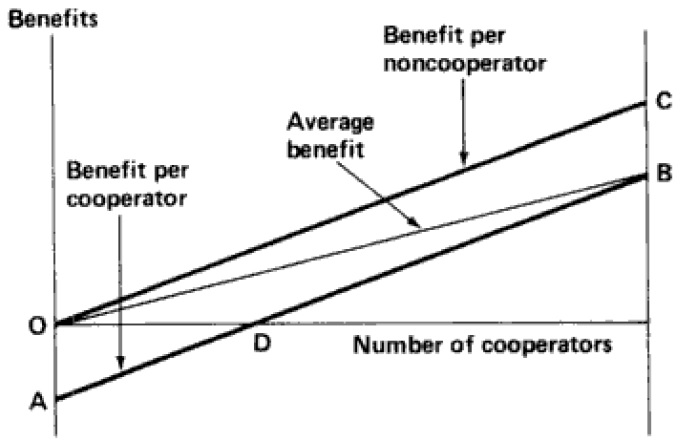}
\caption{{\bf Many-person prisoner's dilemma}. \Didasc Expected benefits as a function of the number of cooperators for the collective group, for the cooperators and for the free riders. Elster (1989).}
\label{Many-person_prisoner's_dilemma}
\end{figure}
\\When the fraction $f$ of taxpayers goes below a certain threshold $f_{th}$, which in Figure~\ref{capital_vs_perc_mat1} (for the adopted values of the free parameters of the model) is slightly less than $40\%$, even being the collective capital $\bar{C}(T)$ still positive, the average capital of taxpayers becomes negative. This means that -- on average -- they pay more than what they receive. In any case, the average social capital of taxtaxpayerspayers is always lower than the average social capital of evaders, i.e. $\bar{C_a}(T)< \bar{C_s}(T) ~\forall f$. Therefore, one should conclude that it is always more convenient, from an individual point of view, to choose a selfish strategy, i.e. to evade taxes, regardless of the fraction $f$ of taxpayers. But we know that the average collective capital $\bar{C}(T)$, for a community of tax evaders only ($f=0\%$), is zero, while the same capital, for a community of taxpayers only ($f=100\%$), reaches its maximum positive value. So, the apparently optimal choice at the individual level leads, at the collective level, towards a sub-optimal results.         
\\This finding is already interesting in itself, since it capture the paradoxical outcome typical of the prisoner's dilemma in the context of game theory. But it is also interesting to notice that, even adopting a minimal number of assumptions, our model gets the same result described by Elster (1989) and shown in Figure~\ref{Many-person_prisoner's_dilemma}. The two heavy lines in the Elster's diagram indicate how the expected benefits, for the cooperators and for the free riders, vary with the number of cooperators (altruists). As in our Figure~\ref{capital_vs_perc_mat1}, the line representing the reward to free riders is constantly above the other one, meaning that noncooperation is individually optimal in terms of selfish benefits. At the same time, it is better for all if all agents cooperate than if nobody does, indeed, $B>0$. The free riders get the largest benefit C, whereas the worst outcome A is reserved for the cooperators. If there are at least D cooperators their benefits become positive. The thin line shows how the average benefits for the collectivity varies with the number of cooperators (also for Elster, by definition, it must begin at 0). The distance between the two heavy lines represents the cost (per altruist) of cooperation. In the figure the cost doesn't vary with the number of cooperators, but in general it may increase or decrease as more people cooperate.
\begin{figure}[tbp]
\centering
\includegraphics[width=\columnwidth]{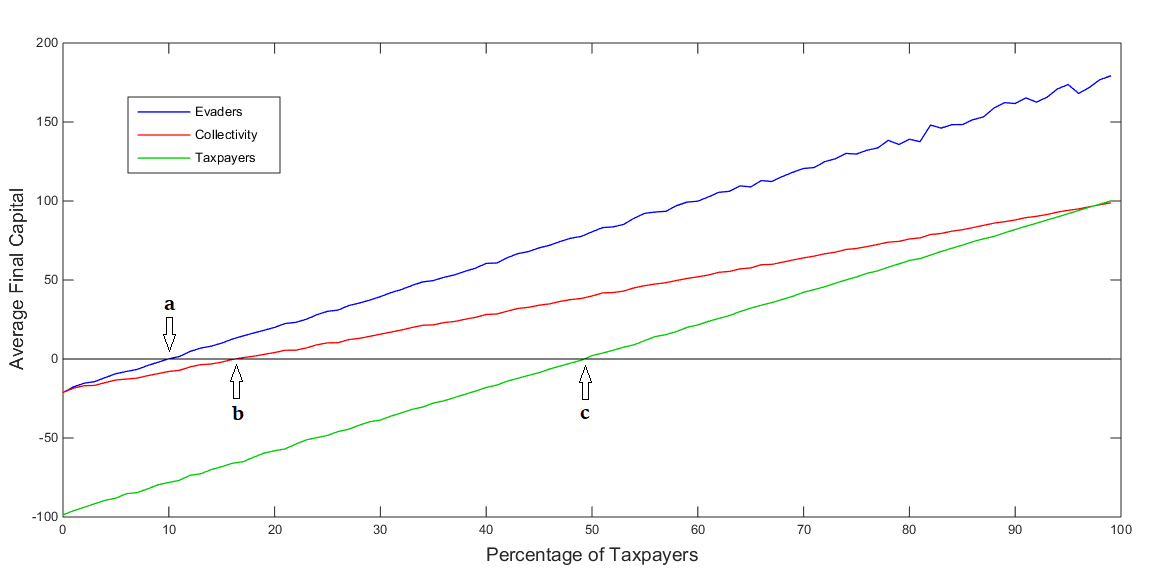}
\caption{{\bf Improved Tax Model}. \Didasc Not rescaled average final capitals $\bar{C}(t)$, $\bar{C_a}(t)$ and $\bar{C_s}(t)$  over $T=100$ turns as a function of the percentage $f$ of taxpayers. All the curves result shifted downward with respect to those in Figure \ref{capital_vs_perc_mat1} and two new thresholds do appear.}
\label{capital_vs_perc_mat2}
\end{figure}
\\Indeed this topic is a collective action dilemma and tax evasion is a problem of the free riders. In individual terms, noncooperation is the most advantageous choice since the capital of tax evaders is always greater than the capital of taxpayers. But the capital of the collectivity increases only thanks to the contribution of altruistic players. Thus, for the collective point of view, groups with more cooperators are favored compared to groups with few cooperators. In fact, altruists pay a cost to obtain benefits for the collectivity, and the more they are, the smaller such a cost is.
\\ However, we could more realistically expect that, if the number of taxpayers is not big enough, selfish tax evaders will get a negative final capital value, because they will need to buy on the market those goods and services which have not been produced due to their evasion. Actually, this is exactly what happens if we do not rescale the curves  shown in Figure~\ref{capital_vs_perc_mat1}. The new results are reported in Figure~\ref{capital_vs_perc_mat2}.
\begin{figure}[tbp]
\centering
\includegraphics[width=0.6\columnwidth]{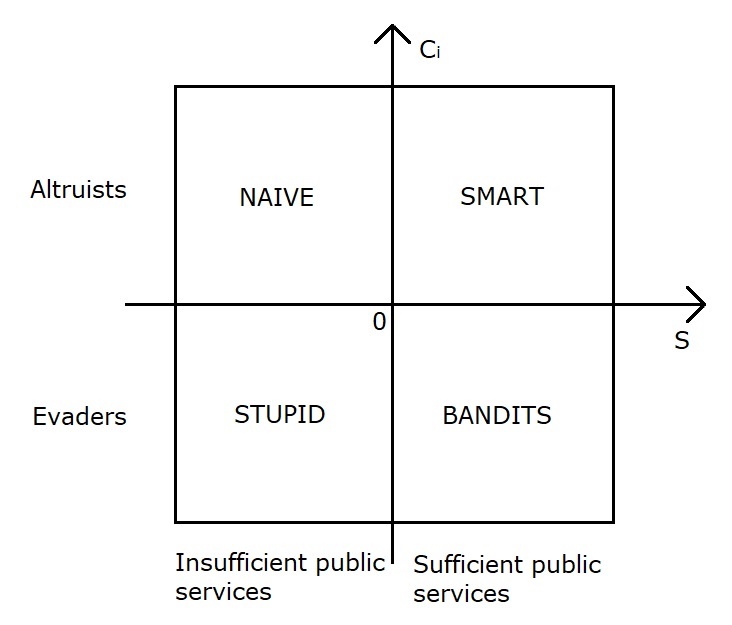} 
\caption{{\bf Adaptation of Cipolla Diagram}. \Didasc Players' behavior sketched as function of both the public services $S$ available for the collectivity and the individual contribution $C_{i}$ of each player to this services. The celebrate four Cipolla's categories can be recognized, strictly depending on the fraction $f$ of taxpayers present in the community.}
\label{cipolla}
\end{figure}
The most evident change with respect to the graph analyzed before is that all the curves have been now shifted downward. As a consequence, we can recognize no longer one but three thresholds, namely $a$, $b$, $c$.
\begin{itemize}
\item If $f<a$, social capital is negative for all of the three categories. As expected, taxes paid by few altruists aren't enough to guarantee the public services for everybody, so evaders damage the community and also themselves. Cooperators suffer twice because they pay taxes and may also need to buy substitutes of public goods. 
\item If $f=a$, tax evaders begin to gain, since there is a sufficient fraction of taxpayers. Taxes paid by altruists suffice to ensure that evaders spend, for substitutes of public goods, less than what they gain (thus making their social capital, on average, positive). Altruists remain in the negative, as the collective social capital.
\item If $f>b$, the average social capital of the collectivity approaches positive values, but still at the expenses of taxpayers whose capital remains, on average, negative.
\item If $f>c$, average capital for taxpayers becomes positive. Above such a threshold, public services are sufficient for everyone (even if the collected resources are less than they should be). Costs are distributed among many altruists and they can finally benefit from their altruistic action. The few evaders still present in the community do continue, of course, to be better off than altruists, since they benefit from the public services without paying taxes and all community members (consciously or unconsciously) settle for a lower quality/quantity of the public service.
\end{itemize}

The scenario just outlined can be effectively summarized by means of an adaptation of the well known Cipolla diagram (Cipolla (1976)), as showed in Figure~\ref{cipolla}. Let's call $S$ the totality of public services available for the collectivity and $C_{i}$ the individual contribution of the each player to this services: $S$ is negative when services are scarce and positive when they are sufficient for guarantee the social wealth, while $C_{i}$ is negative when the agent only benefits from this services without any effort and positive when he contributes to create them. Following these definitions, four quadrants can be clearly recognized in the Cipolla diagram, which identify four types of players depending on the fraction $f$ of taxpayers present in the community:
\begin{enumerate}
\item Smart players: are the taxpayers for $f>c$, since they contribute with their individual capital to ensure a positive level of public services for everybody;
\item Naive players: are the taxpayers for $f<c$, since they maintain positive the collective social capital at their expenses;
\item Bandits: are the tax evaders for $f>a$ since, with their selfish behavior, they get an advantage (free riders) at expenses of the rest of the community;   
\item Stupid players: are the tax evaders for $f<a$ since, with their selfish behavior, not only do penalize the rest of the community, but also themselves.       
\end{enumerate}

\begin{figure}[tbp]
\centering
\includegraphics[width=1\columnwidth]{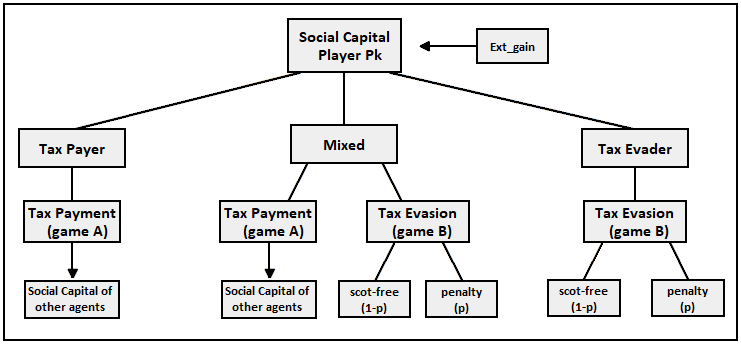}
\caption{{\bf Tax Model with mixed players}. \Didasc Rules of the taxpayersTax evasion model, for a community of $N$ players. Tax evaders play only game B, taxpayers play only game A and mixed players randomly alternate between game~A and game~B.}
\label{Tax_Model_mixed}
\end{figure}

\section{Change of the membership category on complex network}

In this section we introduce the possibility, for both the altruistic and selfish players, to assume an intermediate behavior. For this purpose, we insert a third category between the two ones described in previous sections. Agents belonging to this new category randomly alternate (with probability $0.5$) between game~A and game~B. They will be named "mixed players" because of their dynamic strategy setting, alternatively aimed to preserve their capital (thus evading taxes by game~B), and to cooperate (by game~A), as explained in Figure~\ref{Tax_Model_mixed}.
\\In order to select the transition rules to change category, each player has been given a new variable, named, "believeness" $0<B_i<1$, which specifies the level of commitment in the membership to categories. For both taxpayers and evaders, $B_i=1$ means that the agent is a zealot of her group, whereas the more $B_i$ approaches the zero value, the more it means that the agent is easily influenced and so more inclined to change strategy, becoming a mixed player. For any of the mixed players, the meaning of parameter $B_i$ is slightly different: if $B_i=0.5$, the agent is and remains undecided, whereas values lower that $0.5$ reveal a propensity to become an evader and values greater than $0.5$ show the tendency to become taxpayers.
\\The value of $B_i$ is based on both the imitation of their acquaintances and the individual economic situation.
\\In order to account the first point, we need to introduce an opportune topological structure for our community. In the previous section, the simulations have been performed by assuming a fully connected topology, where each player was able to interact with each other. Now we assume a more realistic social structure, in particular a small-world lattice (Watts and Strogatz 1998), where each player is a node connected with short-range ties (that mimic strong social relationships) to her four neighbors, but with a small rewiring probability ($r=0.02$) of substituting one of those ties with a long-range one (representing a weak social relationship).      
\begin{figure}[tbp]
\centering
\includegraphics[width=\columnwidth]{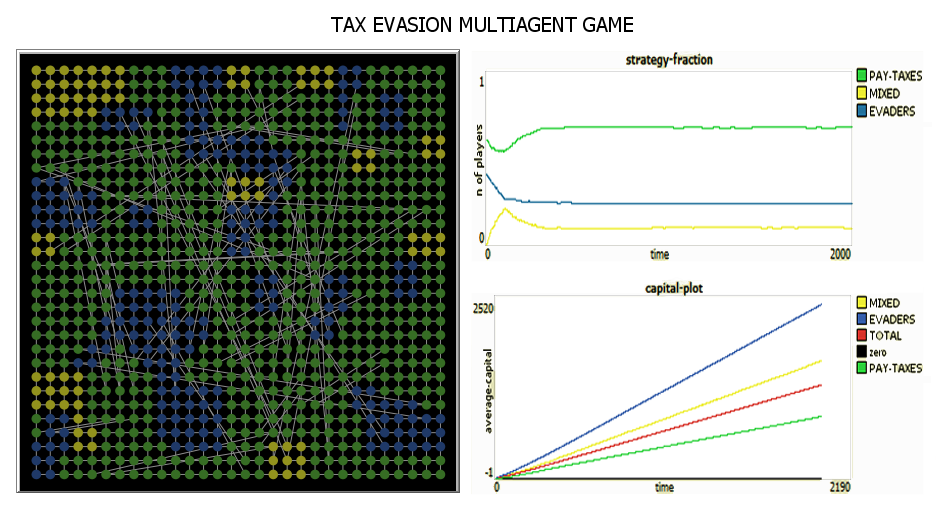}
\caption{{\bf Single run simulation}. \Didasc  An example of the small-world lattice (left panel). Taxpayers are represented with green nodes, evaders with blue nodes and mixed players with yellow ones. The showed configuration of the network has been captured at the end of a single run simulation. The time behavior of the fraction of players belonging to the three considered categories (top right panel) is shown together with that one of the corresponding average social capital (bottom right panel). The initial percentage of taxpayers has been fixed to $60 \%$ (above the critical value for both $IF$ and $CF$ equal to 1).}
\label{schermata-60}
\end{figure}
An example of such a kind of network is reported in the left panel of Figure \ref{schermata-60}, where taxpayers, evaders, and mixed players are represented, respectively, by green, blue, and yellow nodes.
\\For taxpayers and evaders, if at a given time step the number of nearest neighbors belonging to their same category is lower than the sum of the players of the other categories (included the mixed one), the believeness value $B_i$ decreases of a quantity $IF \times \delta B$, where $IF$ is the "Imitation Factor" and $\delta B=0.01$; otherwise, this value increases of the same quantity (of course never exceeding $1$). If, after this, for a certain player of one of these two categories it happens that $B_i \le 0$, that player becomes a mixed player and her new value of $B_i$ is randomly chosen in the interval $[0,1]$. 
\\For mixed players, if the number of nearest neighbors belonging to their same category is lower than the sum of the players of the other two categories, $B_i$ decreases of the quantity $IF \times \delta B$ if the evaders are more than taxpayers, otherwise it increases of the same quantity. If, after this, for a given mixed player it happens that $B_i \le 0$, that player becomes an evader; instead, if for the same player $B_i \ge 1$, she becomes a taxpayer. On the other hand, if the number of mixed players is greater than (or equal to) the sum of the players belonging to the other two categories, $B_i$ moves towards $0.5$ of a quantity $IF \times \delta B$ and the agent maintains her mixed behavior.
\begin{figure}[tbp]
\centering
\includegraphics[width=\columnwidth]{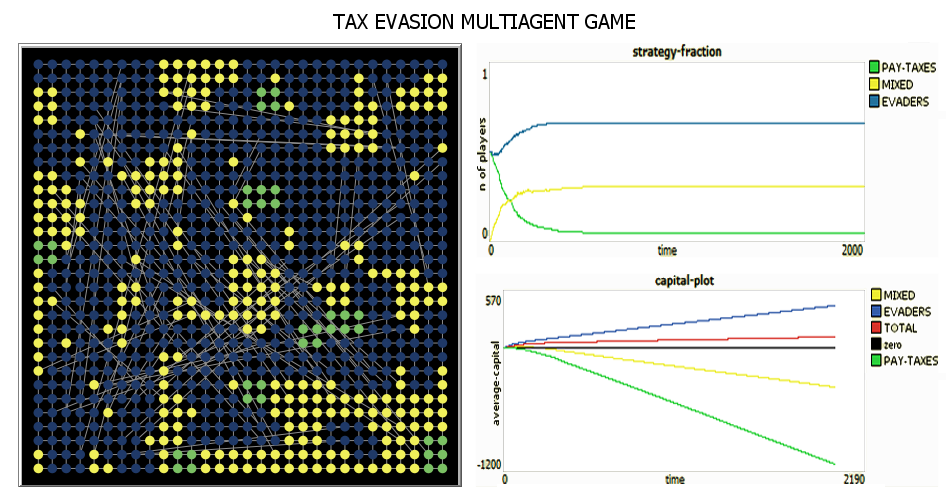}
\caption{{\bf Single run simulation}. \Didasc  Another example of the small-world lattice at the end of another single run simulation, with the same plots of the previous figure but in the case of an initial percentage of taxpayers equal to $50 \%$ (below the critical value for both $IF$ and $CF$ equal to 1).}
\label{schermata-50}
\end{figure}
\\The second mechanism which influences the change of category concerns the economic situation of the players. If the social capital of a given player is negative, the agent will be unsatisfied of her own economic situation and so more prone to change her strategy. So, when their capital is negative, for both evaders and taxpayers the believeness value decreases of a quantity $CF \times \delta B$, where $CF$ is the "Capital Factor"; on the other hand, for mixed players, $B_i$ increases or decreases of the quantity $CF \times \delta B$ depending on their actual state: if $B_i \ge 0.5$ it will increase, if $B_i < 0.5$ it will decrease.
\\Setting to zero the initial percentage of mixed players, in a situation in which the initial percentage of taxpayers is higher than the percentage of evaders, our simulations show that there exist a critical value for the initial percentage of taxpayers (depending on both $IF$ and $CF$), below which the global situation gets always worse and above which gets always better. In Figure~\ref{schermata-60} and in Figure~\ref{schermata-50} are reported the results of two typical single run simulations with $IF=CF=1$ and in correspondence of an initial percentage of taxpayers which is, respectively, above ($60 \%$) and below ($50 \%$) the critical value (that, for these values of $IF$ and $CF$ turns out to be around $55\%$). In the first simulation, the final economic condition appears to be good for all the categories and in the community there is a majority of players who pay taxes; instead, in the second one, the final economic situation is good only for the evaders and these players represent the majority of the agents. 
\begin{figure}[tbp]
\centering
\includegraphics[width=0.95\columnwidth]{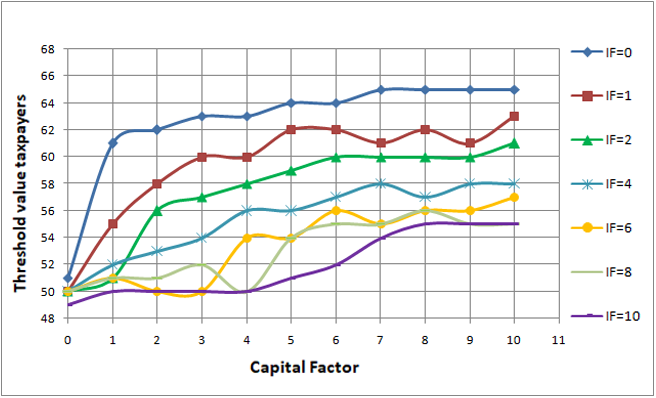}
\caption{{\bf Thresholds diagram}. \Didasc  Threshold values for the initial percentage of taxpayers as function of the capital factor $CF$ and for different values of the imitation factor $IF$.}
\label{CF_IF}
\end{figure}
\\The threshold values for the initial percentage of taxpayers as a function of $CF$ are showed in Figure~\ref{CF_IF} for different values of $IF$. As we can see, for a given value of $IF$, generally the critical value of the initial percentage of pay taxes rapidly increases with $CF$, then it tends to oscillate around a stationary asymptotic value which decreases with $IF$. For $IF=0$, i.e. without imitation, a change in strategy is due to $CF$ only: when $CF$ is low, i.e. when the dissatisfaction for a negative economic situation is not significant,  a small initial percentage of taxpayers is enough to induce a final positive trend for the whole community; on the contrary, when $CF$ is high, i.e. when a negative capital heavily acts on the personal dissatisfaction, the initial percentage of taxpayers has to be more consistent in order to counterbalance the tendency of the altruistic component to be penalized with respect to the selfish one. Such an effect is reduced by the presence of imitation, which in average helps the initial fraction of taxpayers -- that should be, in any case, always greater than $50\%$ -- to spread around the community. 
\\Following such results, a reasonable policy implication could be to induce a sense of satisfaction in taxpayers, in order to reduce the temptation to evade, even when the personal economic situation is bad. Thus, it should be better, for the Government, to care of taxpayers than of the evaders: for instance, an educational policy spreading a tax morale is expected to be more effective than a tax amnesty, because it operates in such a way that individuals feel themselves rewarded by institutions. This can also impact on the number of taxpayers, which has been described as a key factor in determining the average social capital.

\section{Variation of the tax, penalty and audit probability}

\begin{figure}[tbp]
\centering
\includegraphics[width=1.0\columnwidth]{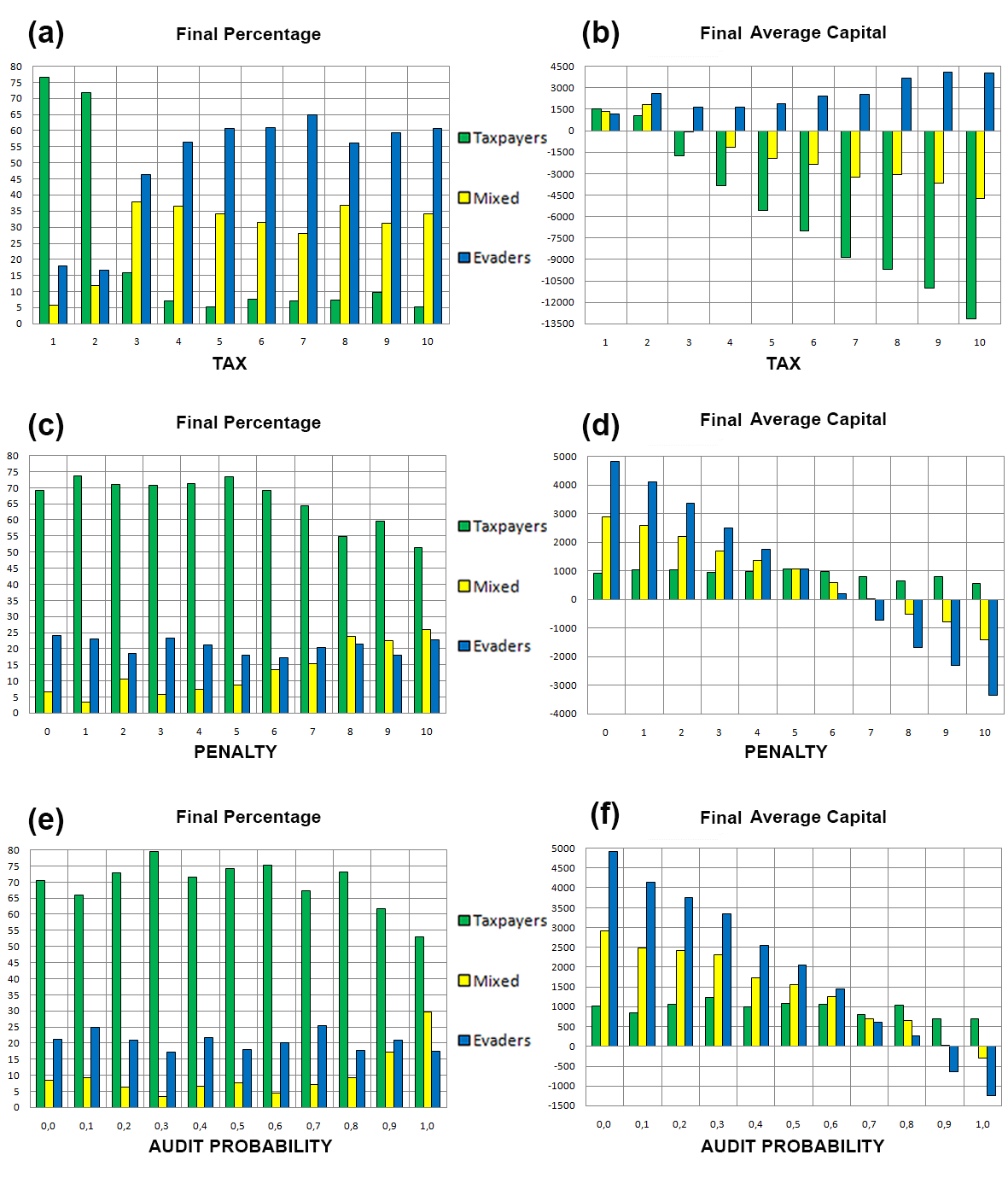} 
\caption{\Didasc Final percentage and final average capital after 2000 turns of the three categories for increasing values of, respectively, tax (a-b), penalty (c-d) and audit probability (e-f). The initial percentage of taxpayers is always equal to 60\%. }
\label{percentage-and-capital}
\end{figure}

In this last section we want to analyze the evolution of the system as function of the three main free parameters of our model, namely, the tax, the penalty and the probability of an audit. It is, indeed, worth to notice that changing the value of the external gain $g$ would produce only a symmetric rescaling effect on the final capital of all the social components. Therefore, we neglect its variations and let $g=1$, without loss of generality. In previous sections, parameters values were: $d=2$, $h=3$, and $p=0.4$. In all simulations we also fix $IF$ and $CF$ to 1, and we start with two different initial percentage of taxpayers, above and below the  $55\%$ threshold (which has already been shown in Figure~\ref{CF_IF}).
\\In the first setting, taxpayers are $60\%$, mixed players $0\%$ and evaders $40\%$. In Figure~\ref{percentage-and-capital} the final percentage of agents (panel a) and the final average capital (panel b) are reported for the three categories and after 2000 turns, in correspondence of increasing values of the tax $d$ (we verified that 2000 turns are sufficient to stabilize the final composition of the population). It clearly appears that for $d<3$ units (i.e. when the tax is lower than the penalty $h$) the final composition of the population is dominated by taxpayers, with percentage above 70 \%, and there is a corresponding good economical situation for all the three categories. For higher values of the tax (i.e. for $d \ge h$), the evolution of the system leads to a majority of tax evaders and this situation fits good only to evaders. This means that, as one could expect, keeping the amount of taxes below the penalty induces tax payment and results in a decrease of the evasion. Figure~\ref{percentage-and-capital} also shows the final percentage and the final average capital for the three categories and after 2000 turns, but for increasing values of the penalty $h$ (panels c and d) and the audit probability $p$ (panels e and f), respectively. Looking at the average capital in panels (d) and (f), we can see that for values of the penalty $h>6$ and for the audit probability $p>0.8$, the economic situation of the evaders becomes worse than the ones of taxpayers and mixed players. But, curiously, this does not induce necessarily to a reduction of tax evasion: in fact, panels (c) and (e) show that the final percentage of evaders does not result to be lower for $h>6$ and $p>0.8$. On the contrary, above such thresholds, the number of mixed players increases at expenses of taxpayers and the final percentage of taxpayers tends to go below the initial one. Of course, it would be interesting to test the occurrence of such a paradoxical effect in the real world. 
\begin{figure}[tbp]
\centering
\includegraphics[width=1.0\columnwidth]{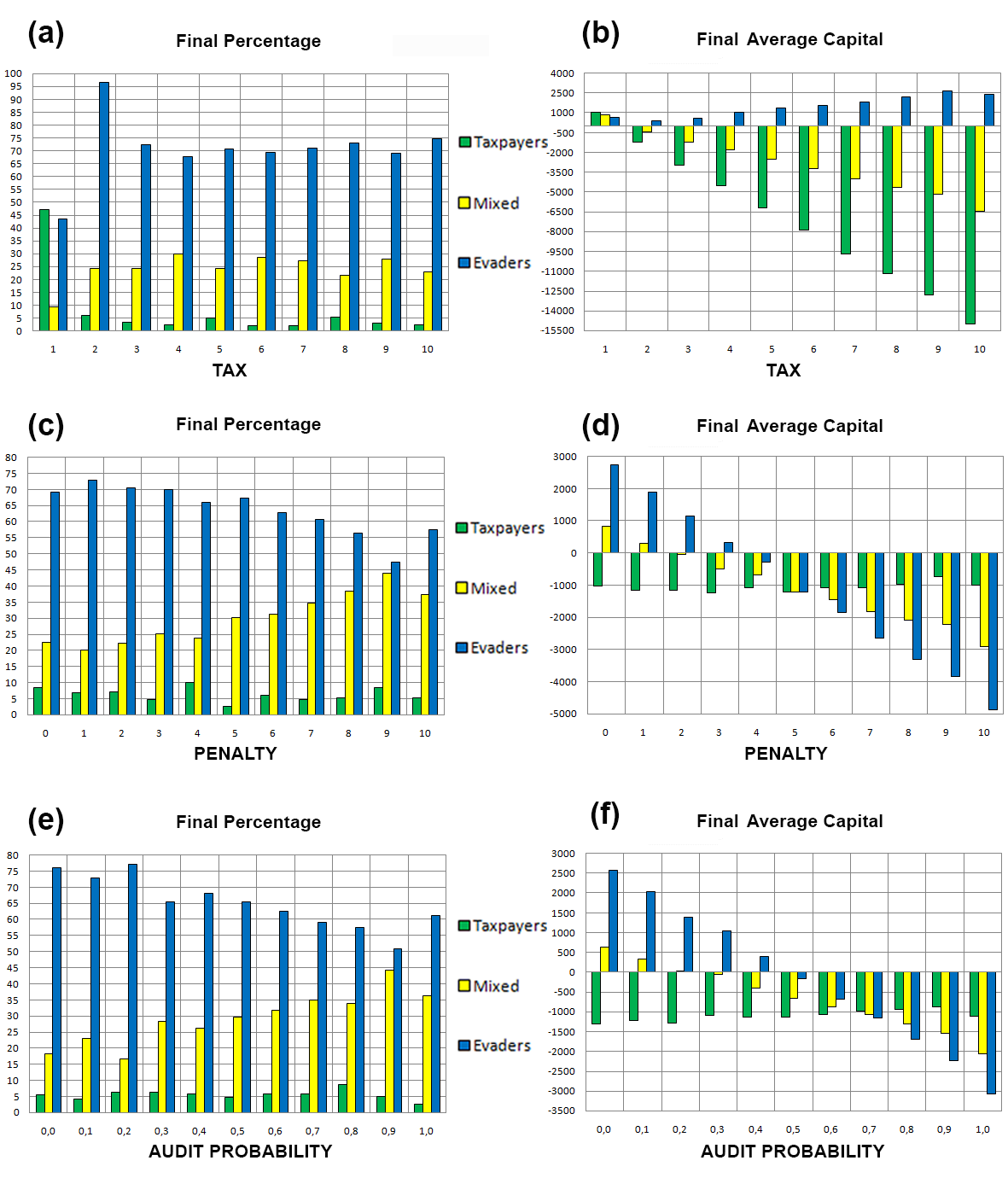} 
\caption{\Didasc Final percentage and final average capital after 2000 turns of the three categories for increasing values of, respectively, tax (a-b), penalty (c-d) and audit probability (e-f). The initial percentage of taxpayers is always equal to 50\%. }
\label{percentage-and-capital2}
\end{figure}
\\Finally, let us address the second case, with a percentage of taxpayers fixed to 50\% (below the threshold), mixed players to 0\% and evaders to 50\%. Figure~\ref{percentage-and-capital2} shows the results for the variation of tax (panels a and b), penalty (panels c and d) and audit probability (panels e and f). In this case, as expected, the final capital of the taxpayers is always lower compared to the case above the threshold, but looking at the final capital of the evaders (right panels) we can draw same conclusions as in the previous case: small values for the tax and high values for both the penalty and the audit probability do induce a worse economic situation for the evaders. On the other hand, the final composition of the community (left panels) is always dominated by the component of tax evaders, apart from the case with $d=1$. However, differently from the previous case, an increase of either the penalty or the audit probability, has now the beneficial effect of reducing the number of evaders, who tend to change strategy becoming mixed players. It is interesting to notice that, in correspondence of the maximum values $h=10$ and $p=1.0$, this trend is inverted since the percentage of evaders starts again to slightly increase.

\section*{Conclusions}
\addcontentsline{toc}{section}{Conclusions}

In this paper we presented a very simple toy model of tax evasion in order to explore, by means of extended agent-based simulations, under which conditions people behave altruistically, by paying taxes, or, on the contrary, behave selfishly, by evading taxes.
\\In the first part of the paper, the impact of a varying fraction of altruistic players on the final capital of a fully connected community has been shown. Surprisingly, the simulation results showed that, despite its simplicity, the model is able to capture the real trend of the expected benefits for a collective group in presence of a variable number of cooperators and free riders, as already described by Elster (1989). A further improvement of the model, which accounts for the need of substitutes to non-produced public goods (because of high levels of tax evasion), has allowed to identify a threshold, in the fraction of taxpayers, below which evaders not only create a damage for the collectivity but also for themselves (thus falling in the category of "stupid people" in the famous 1976 Cipolla's diagram).
\\In the second part of the paper, the model was enriched by the introduction of some extensions: a small-world network topology for the social community (driving the imitation), and a third category of "mixed" players (sometimes altruist, sometimes selfish). New interesting results have been obtained, showing the presence of a threshold, in the initial percentage of taxpayers, able to ensure an average economic advantage to this category of players at the end of the simulations. Such a threshold is influenced by the individual propensity of agents to imitate and by their sensitivity with respect to their personal economic situation. Finally, an extended parametric study has shown how the amount of taxes to pay, the penalty for evaders and the audit probability, do influence both the final composition of the considered community and the final average capital of its three social components (taxpayers, evaders and mixed players).

\section*{Acknowledgments}
Authors thank Andrea Rapisarda for useful discussions.

\section*{References}

Allingham, M.G., \& Sandmo A. (1972). Income Tax Evasion: A Theoretical Analysis. \textit{Journal of Public Economics, vol.1}, 323-338.

Alm, J. (2012). Measuring, explaining, and controlling tax evasion: lessons from theory, experiments, and  eld studies. \textit{International Tax and Public Finance, 19 (1)}, 54–77.

Alm, J., McClelland, G.H., \& Schulze, W.D. (1992). Why do people pay taxes? \textit{Journal of Public Economics, vol.48}, pp.21–38.

Alstads{\ae}ter A., Johannesen, N., \& Zucman, G. (2017). Tax Evasion and Inequality, \textit{NBER Working Paper No. 23772}, September 2017. Invited resubmission: \textit{American Economic Review}.

Andreoni, J., Erard, B., \& Feinstein, J. (1998). Tax compliance. \textit{Journal of Economic Literature, 36 (2)}, 818–860.

Baumol, W. (1952). Welfare Economics and the Theory of the State. \textit{Cambridge, Massachusetts: Harvard University Press}.

Bazart, C., Bonein, A., Hokamp, S., \& Seibold, G. (2016). Behavioural economics and tax evasion–calibrating an agent-based econophysics model with experimental tax compliance data. \textit{Journal of Tax Administration, vol.2 (1)}, pp.126–144.

Becker, G.S. (1968). Crime and punishment: an economic approach.\textit{ The Journal of Political Economy, vol.76 (2)}, pp.169–217.

Bertotti, M.L. \& Modanese, G. (2014). Micro to macro models for income distribution in the absence and in presence of tax evasion. \textit{Applied Mathematics and Computation, vol.224}, pp.836–846.

Bertotti, M.L. \& Modanese, G. (2016). Microscope Models for the Study of Taxpayer Audit Effects, \textit{arXiv} 1602.08467v1 [q- n.GN] 18 February 2016.

Biondo, A.E., Pluchino, A., Rapisarda, A., \& Helbing, D. (2013a). Reducing financial avalanches by random investments. \textit{Phys. Rev. E} 88(6):062814.

Biondo, A.E., Pluchino, A., Rapisarda, A., \& Helbing, D. (2013b). Are random trading strategies more successful than technical ones. \textit{PLoS One} 8(7):e68344.

Biondo, A.E., Pluchino, A., \& Rapisarda, A. (2013c). The beneficial role of random strategies in social and financial systems. \textit{J. Stat. Phys}. 151(3-4):607-622.

Biondo, A.E., Pluchino, A., \& Rapisarda, A. (2014). Micro and macro benefits of random investments in financial markets. \textit{Cont. Phys.} 55(4):318-334.

Biondo, A.E., Pluchino, A., \& Rapisarda, A. (2015). Modeling financial markets by self-organized criticality. \textit{Phys. Rev. E} 92(4):042814.

Biondo, A.E., Pluchino, A., \& Rapisarda, A. (2017). Contagion dynamics in a multilayer network model of financial markets. \textit{Italian Economic Journal}, DOI 10.1007/s40797-017-0052-4. 

Bloomquist, K.M. (2006). A comparison of agent-based models of income tax evasion. \textit{Social Science Computer Review, vol.24 (4)}, pp.411–425.

Buehn, A. \& Schneider, F. (2012). Shadow economies around the world: novel insights, accepted knowledge, and new estimates. \textit{International Tax and Public Finance, vol.19 (1)}, pp.139–171.

Callen, E., \& Shapero, D. (1974). A theory of social imitation. \textit{Physics Today, vol. 27(7)}. DOI: 10.1063/1.3128690

Cipolla, C. M. (1976). The Basic Laws of Human Stupidity. \textit{The Mad Millers}.

Clotfelter, C.T. (1983). Tax Evasion and Tax Rates: An analysis of individual returns, \textit{The Review of Economics and Statistics vol.65},  pp.363-73.

Crane, S.E. \& Nourzad, F. (1987). On the Treatment of Income Tax Rates in Empirical Analysis of Tax Evasion, \textit{Kyklos, vol.40}, pp.338-348.
 
Dalamagas, B. (2011). A Dynamic Approach to Tax Evasion, \textit{Public Finance Review, vol.39 (2)}, pp.309-326.

Elsenbroich C., \& Gilbert N. (2014). Imitation and Social Norms. In: \textit{Modelling Norms}. Springer, Dordrecht.

Elster, J. (1989). Nuts and bolts for the social sciences, \textit{Cambridge University Press}.

Epstein, R.A. (1993). Altruism: Universal and Selective, \textit{Social Service Review, vol.67, (3)}, pp.388-405

Feld, L.P., \& Frey, B.S. (2007). Tax Compliance as the Result of a Psychological Tax Contract: The Role of Incentives and Responsive Regulation, \textit{Law \& Policy}, doi.org/10.1111/j.1467-9930.2007.00248.x 

Fortin, B., Lacroix, G., \& Villeval, M.C. (2007). Tax evasion and social interactions. \textit{Journal of Public Economics, vol.91}, pp.2089–2112.

Hardin, R. (1995). One for All: The Logic of Group Conflict. \textit{Princeton University Press, Princeton, NJ}.

Heckathorn, D.D. (1996). The dynamics and dilemmas of collective action. \textit{American Sociological Review, vol.61(2)}, pp.250–277.

Hokamp, S. (2013). Income tax evasion and public goods provision – theoretical aspects and agent-based simulations. \textit{PhD thesis. Brandenburg University of Technology Cottbus}.

Hokamp, S. \& Pickhardt, M. (2010). Income tax evasion in a society of heterogeneous agents – evidence from an agent-based model. \textit{International Economic Journal, 24 (4)}, 541–553.

Internal Revenue Service (2016). IRS Tax Gap Estimates for Tax Years 2008–2010, \textit{https://www.irs.gov/PUP/newsroom/tax gap estimates for 2008 through 2010.pdf} (accessed 23 November 2016).

Kirchler, E. (2007). The Economic Psychology of Tax Behavior, \textit{Cambridge University Press, Cambridge}.

La Porta, R., Lopez-de-Salanes, F ., Shleifer, A., \& Vishny, R. (1999). The Quality of Government. \textit{Journal of Law, Economics, and Organization, vol.15 (1)}, pp. 222–279. doi:10.1093/jleo/15.1.222.

Lima, F.W.S. \& Zaklan, G. (2008). A multi-agent-based approach to tax morale. \textit{International Journal of Modern Physics C: Computational Physics and Physical Computation, vol.19 (12)}, pp.1797–1808.

McDonald, R.I., \& Crandall C.S. (2015). Social norms and social influence, \textit{Current Opinion in Behavioral Sciences, vol.3}, pp.147–151.

Myles, G.D. \& Naylor, R.A. (1996). A Model of Tax Evasion with Group Conformity and Social Custom. \textit{European Journal of Political Economy, vol.12 (1)}, pp.49-66.

Nicolaides P. (2014). Tax Compliance Social Norms and Institutional Quality: An Evolutionary Theory of Public Good Provision. \textit{Taxation Papers 46}, Directorate General Taxation and Customs Union, European Commission.

Oates, L. (2015). Review of recent literature. \textit{Journal of Tax Administration, vol.1 (1)}, pp.141–150.

Panades, J. (2004). Tax Evasion and Relative Tax Contribution,  \textit{Public Finance Review, vol.32 (2)}, pp.183-195.

Pickhardt, M. \& Prinz, A. (2014). Behavioral dynamics of tax evasion – a survey. \textit{Journal of Economic Psychology, vol.40}, pp.1–19.

Pluchino, A., Rapisarda, A., \& Garofalo C. (2010). The Peter principle revisited: A computational study. \textit{Physica A: Statistical Mechanics and its Applications}, 389(3):467-472.

Pluchino, A., Rapisarda, A., \& Garofalo C. (2011). Efficient promotion strategies in hierarchical organizations. \textit{Physica A: Statistical Mechanics and its Applications}, 390(20):3496-3511.

Pluchino, A., Biondo, A. E., \& Rapisarda A. (2018). Talent versus Luck: the role of randomness in success and failure. \textit{Advances in Complex Systems}, Vol.21 No.03n04, 1850014.

Poterba J.M. (1987). Tax Evasion and Capital Gains Taxation. \textit{American Economic Review, vol.77}, pp.234-239.

Rapoport, A. (1974). Prisoner’s dilemma- Recollections and observations. In \textit{Game Theory as a Theory of Conflict Resolution} (ed. Rapoport A). Reidel, Dordrecht, The Netherlands, pp. 17–34.
 
Robbins, L. (1932). An Essay on the Nature and Significance of Economic Science. \textit{Macmillan, London}.

Slemrod, J. (2007). Cheating ourselves: the economics of tax evasion. \textit{Journal of Economic Perspectives, 21 (1)}, 25–48.

Slemrod, J., \& Yitzhaki, S. (2002). Tax avoidance, evasion and administration. \textit{Handbook of Public Economics, North-Holland, Amsterdam}, pp. 1425–1470.

Srinivasan, T.N. (1973). Tax evasion: a model. \textit{Journal of Public Economics, vol.2}, pp.339-346.

Stevens, J.B. (2018). The Economics Of Collective Choice. \textit{Routledge 2018, NY}.

Torgler, B. (2002). Speaking to theorists and searching for facts: tax morale and tax compliance in experiments. \textit{Journal of Economic Surveys, vol.16}, pp.657–683.

Torgler, B., \& Schneider, F. (2009). The impact of tax morale and institutional quality on the shadow economy. \textit{Journal of Economic Psychology, vol.30, (2)}, pp.228-245, doi.org/10.1016/j.joep.2008.08.004.

Traxler C. (2006). Social Norms and Conditional Cooperative Taxpayers. \textit{Department of Economics Discussion Paper No. 2006-28}. Munich: University of Munich.

Vale, R. (2015). A Model for Tax Evasion with Some Realistic Properties. \textit{arXiv:1508.02476 [q-fin.EC]}

Voss, T. (2001). Game-theoretical perspectives on the emergence of social norms. \textit{ Social Norms (eds Hechter M and Opp KD)}. Russell Sage Foundation, New York, pp.105–136.

Watts, D. J. and Strogatz, S. H. (1998). Collective dynamics of 'small-world' networks. \textit{Nature} (London) 393, 440.

Yaniv, G. (2013). Tax Evasion, Conspicuous Consumption and the Income Tax Rate. \textit{Public Finance Review, vol.41 (3)}, pp.302-316
 
Yitzhaki S. (1974). A note on Income Tax Evasion: A Theoretical Analysis. \textit{Journal of Public Economics, vol.3}, 201-202.

Yitzhaki, S. (1987). On the excess burden of tax evasion. \textit{Public Finance Quaterly, vol.15}, pp.123-37.

Zaklan, G., Lima, F.W.S., \& Westerhoff, F. (2008). Controlling tax evasion  fluctuations. \textit{Physica A: Statistical Mechanics and its Applications, vol.387 (23)}, pp.5857–5861.

Zappal\`a, D.A., Pluchino, A., \& Rapisarda, A. (2014). Selective altruism in collective games. \textit{Physica A, vol.410}, pp.496-512.

\end{document}